\documentclass{amsproc}

\theoremstyle{definition}

\theoremstyle{remark}

\numberwithin{equation}{section}

\newcommand{\rz}{\mathbb{R}}       % Relle Zahlen

\begin{document}

\title[Isoperimetric problem for ovals on the plane]
{Connection between the Lieb--Thirring conjecture for Schr\"odinger operators
and an isoperimetric problem for ovals on the plane}
%    Information for first author
\author{Rafael D. Benguria}
%    Address of record for the research reported here
\address{Department of Physics, P. Universidad Cat\'olica de Chile, 
Casilla 306, Santiago 22, Chile}
%    Current address
\email{rbenguri@fis.puc.cl}
%    \thanks will become a 1st page footnote.
\thanks{This work  was supported in part by Fondecyt (Chile), 
Projects \# 102--0844 and \# 702--0844}
\author{Michael Loss}
%    Address of record for the research reported here
\address{School of Mathematics, Georgia Institute of Technology}
%    Current address
\email{loss@math.gatech.edu}
%    \thanks will become a 1st page footnote.
\thanks{Work partially supported by NSF Grant DMS 03--00349}

%    General info
\date{}


\begin{abstract}
To determine the sharp constants for the one dimensional Lieb--Thirring 
inequalities with exponent $\gamma \in (1/2,3/2)$ is still an open problem. 
According to a conjecture by Lieb and Thirring the sharp constant for these 
exponents should be attained by potentials having only one
bound state. Here we exhibit a connection between the Lieb--Thirring 
conjecture for $\gamma=1$ and an isporimetric inequality for ovals in the 
plane. 
\end{abstract}

\maketitle

\section{Introduction}
The Lieb--Thirring inequalities are one of the main tools in the proof of the 
stability of matter 
\cite{LiTh75} (see also the review article \cite{Li76} or \cite{Li00}). 
Let $H=-\Delta + V$ be the Schr\"odinger operator acting on 
$L^2(\rz^n)$, $n \ge 1$ and 
denote by $e_1 \le e_2 \le \dots < 0$ the negative eigenvalues of $H$. The 
Lieb--Thirring inequalities are given by
\begin{equation}
\sum_{j \ge 1} {\vert e_j \vert}^{\gamma} \le L_{\gamma,n} \int_{\rz^n} 
V_{-}(x)^{\gamma+n/2} \, dx,
\label{eq:1.1}
\end{equation}
where $V_{-}(x) \equiv \max(-V(x),0)$ is the negative part of the potential. 
The above inequalities hold for $\gamma \ge 1/2$ when $n=1$, for $\gamma >0$ 
when $n=2$, and for $\gamma \ge 0$ for $n \ge 3$. The case $\gamma=1/2$, 
$n=1$ was established by T. Weidl \cite{We96}. The case $\gamma=0$, 
$n \ge 3$ was established independently by M. Cwikel, E.H. Lieb and 
G.V. Rosenbljum. One can show in general that 
$L_{\gamma,n} \ge L_{\gamma,n}^c$, where
\begin{equation}
L_{\gamma,n}^c = 2^{-n} \pi^{-n/2} \frac{\Gamma(\gamma+1)}
{\Gamma(\gamma+1+n/2)}
\label{eq:1.2}
\end{equation}
are the semiclassical constants. Define 
$R_{\gamma,n} \equiv L_{\gamma,n}/ L_{\gamma,n}^c \ge 1$. 
Aizenman and Lieb proved that $R_{\gamma,n}$ decreases as 
$\gamma$ increases \cite{AiLi78}. In \cite{LiTh76} it is proven that 
$L_{3/2,1} = L_{3/2,1}^c$ and thus,  
$L_{\gamma,1} = L_{\gamma,1}^c$, for all $\gamma \ge 3/2$. 
For $n>1$, Laptev and Weidl \cite{LaWe00a} proved 
$L_{3/2,n} = L_{3/2,n}^c$ hence, 
$L_{\gamma,n} = L_{\gamma,n}^c$, for all $\gamma \ge 3/2$.
The sharp constant for $\gamma=1/2$ and $n=1$, $L_{1/2,1}=1/2$ was proved in 
\cite{HuLiTh98}.
For best constants up to date see \cite{HuLaWe00}. 

For $n=1$, the sharp constants $L_{\gamma,1}$ are not known for values of 
$\gamma$ in the interval $(1/2,3/2)$. However, in 1976 Lieb and Thirring 
\cite{LiTh76} conjectured that the sharp constants are attained for 
potentials that have only one bound state, and therefore
\begin{equation}
L_{\gamma,1} \equiv L_{\gamma , 1}^1 = \frac{1}{\sqrt{\pi}} 
\frac{1}{\gamma-1/2} 
\frac{\Gamma(\gamma+1)}{\Gamma(\gamma+1/2)}
\left(\frac{\gamma-1/2}{\gamma+1/2} \right)^{\gamma+1/2}.
\label{eq:1.3}
\end{equation}

In this manuscript we establish a connection between the 
Lieb--Thirring conjecture for $\gamma=1$ and $n=1$ and an 
isoperimetric inequality for closed curves in the plane which are 
smooth, have  positive curvature and length $2 \pi$.  
The rest of the article is organized as follows. 
In Section 2, we provide a new and direct method for maximizing the 
lowest eigenvalue of one dimensional Schr\"odinger operators. 
In Section 3 we establish the aforementioned connection with a problem 
for closed curves in the plane. We should emphasize that the 
isoperimetric problem that we allude to is also still open.

\section{Maximizing the first eigenvalue}

The problem of maximizing the lowest eigenvalue of the one-dimensional 
Schr\"odinger operator on the line subject to a constraint on integrals of 
powers of the potential was first considered by Joseph Keller in 1961 
\cite{Ke61}. See also \cite{LiTh76,AsHa87,BeVe92,Ve02}. 

Consider the Schr\"odinger operator,
\begin{equation}
H=-\frac{d^2}{dx^2} + V
\label{eq:2.1}
\end{equation}
defined on $L^2(\rz)$, and let $-\lambda_1$ be the lowest eigenvalue. 
Then,
\begin{equation}
\lambda_1^{\gamma} \le L_{\gamma , 1}^1 \int_{-\infty}^{\infty} 
V_{-}(x)^{\gamma+1/2} \, dx,
\label{eq:2.2}
\end{equation}
for all $\gamma>1/2$, where the sharp constants $L_{\gamma,1}^1$ are 
given by,
\begin{equation}
L_{\gamma,1}^1 = \frac{1}{\sqrt{\pi}} \frac{1}{\gamma-1/2} 
\frac{\Gamma(\gamma+1)}{\Gamma(\gamma+1/2)}
\left(\frac{\gamma-1/2}{\gamma+1/2} \right)^{\gamma+1/2}.
\label{eq:2.3}
\end{equation}
Keller's proof uses the Direct Calculus of Variations. When the exponent 
$\gamma=1$ there is a very simple argument to compute the best constant. 
We give the full argument in the sequel, because it is important in our 
later derivation of the connection between the Lieb--Thirring conjecture and 
an isoperimetric inequality for ovals in $\rz^2$. 

Let $u_1$ and $-\lambda_1$ be the normalized ground state and the lowest 
eigenvalue of the Schr\"odinger operator $H=-d^2/dx^2 - V$ on $L^2(\rz)$, 
where $V \ge 0$. Thus,
\begin{equation}
-u_1''-V \, u_1 = - \lambda_1 u_1, \qquad \mbox{in $\rz$}.
\label{eq:2.4}
\end{equation}
Multiplying (\ref{eq:2.4}) by $u_1$ and integrating in $\rz$, we get
\begin{equation}
\lambda_1 = \int_{\rz} V \, u_1^2 \, dx - \int_{\rz} (u_1')^2 \, dx.
\label{eq:2.5}
\end{equation}
Since,
$$
V \, u_1^2 \le K \, V^{3/2} + \frac{4}{27 \, K^2} u_1^6,
$$
for all $K >0$, from (\ref{eq:2.5})
we get,
\begin{equation}
\lambda_1 \le  K \int_{\rz} V^{3/2} \, dx + \frac{4}{27 \, K^2} 
\int_{\rz} u_1^6 \, dx - \int_{\rz} (u_1')^2 \, dx. 
\label{eq:2.6}
\end{equation}
However, if $\int_{\rz} u_1^2 \, dx=1$, 
\begin{equation}
\int_{\rz} (u_1')^2 \, dx \ge \frac{\pi^2}{4} \int_{\rz} u_1^6 \, dx,
\label{eq:2.7}
\end{equation}
so, choosing $K=4/(3 \sqrt{3} \pi)$, we finally get
\begin{equation}
\lambda_1 \le \frac{4}{3 \sqrt{3} \pi} \int_{\rz} V^{3/2} \, dx = L_{1,1}^1 
\int_{\rz} V^{3/2} \, dx
\label{eq:2.8}
\end{equation}
which is Keller's result for $\gamma=1$.
For completeness we give an elementary proof of (\ref{eq:2.7}). 
First make the change of variables $x \to s$ given by
\begin{equation}
s = \int_{-\infty}^{x} u_1^2 \, dy.
\label{eq:2.9}
\end{equation}
Here, $s: 0 \to 1$, and $ds/dx=u_1^2$. With this change of variables we have,
\begin{equation}
\int_{-\infty}^{\infty} u_1^6 \, dx = \int_0^1 u_1^4 \, ds, 
\label{eq:2.10}
\end{equation}
and
\begin{equation}
\int_{-\infty}^{\infty} (u_1')^2 \, dx = \int_0^1 (\dot u_1)^2 u_1^2 \, ds, 
\label{eq:2.11}
\end{equation}
where $\dot u_1 \equiv d u_1/ds$. Since $u_1$ goes to zero at 
$x=\pm \infty$ we have
$u_1(s=0)=u_1(s=1)=0$. Finally, if we call $w \equiv u_1^2$, 
$\int_{-\infty}^{\infty} u_1^6 = \int_0^1 w^2 \, ds$ and 
$\int_{-\infty}^{\infty} (u_1')^2 \, dx 
=(1/4) \int_0^1 {\dot w}^2 \, ds$. In terms of $w(s)$, (\ref{eq:2.7}) 
is given by
\begin{equation}
\int_0^1 {\dot w}^2 \, ds \ge \pi^2 \int_0^1 w^2 \, ds,
\label{eq:2.12}
\end{equation}
which follows from the fact that the first Dirichlet eigenvalue of the 
interval $(0,1)$ is $\pi^2$. One can obtain the cases with $\gamma \neq 1$ 
in (\ref{eq:2.2}) in a similar way (see the Appendix).

\section{Maximizing the sum of the first two eigenvalues and the connection 
with a geometric problem in $\rz^2$.}

Consider the Schr\"odinger operator
$$
H=-\frac{d^2}{dx^2} - V,
$$
on $L^2(\rz)$ with $V \ge 0$ such that $\int V^{3/2} \, dx < \infty$. 
Assume $H$ has at least two negative eigenvalues, and denote by $-\lambda_1$
and $-\lambda_2$ the lowest two eigenvalues and $u_1$, $u_2$ the corresponding
normalized eigenfunctions. As before, we have
\begin{equation}
\lambda_1 = \int_{\rz} V \, u_1^2 \, dx - \int_{\rz} (u_1')^2 \, dx,
\label{eq:3.1}
\end{equation}
and
\begin{equation}
\lambda_2 = \int_{\rz} V \, u_2^2 \, dx - \int_{\rz} (u_2')^2 \, dx.
\label{eq:3.2}
\end{equation}
Adding these two equations and using the pointwise bound,
$$
V (u_1^2+u_2^2)  \le K \, V^{3/2} + \frac{4}{27 \, K^2} (u_1^2+u_2^2)^3,
$$
we get 
\begin{equation}
\lambda_1+\lambda_2 \le  K \int_{\rz} V^{3/2} \, dx + 
\frac{4}{27 \, K^2} \int_{\rz} 
(u_1^2+u_2^2)^3 \, dx
- \int_{\rz} \left((u_1')^2+(u_2')^2 \right)\, dx. 
\label{eq:3.3}
\end{equation}
In order to prove the Lieb--Thirring conjecture for $\gamma=1$ in the 
special case of
potentials having only two eigenvalues, it would be enough to prove
\begin{equation}
\int_{\rz} \left((u_1')^2+(u_2')^2 \right)\, dx \ge \frac{\pi^2}{4}
\int_{\rz} (u_1^2+u_2^2)^3 \, dx,
\label{eq:3.4}
\end{equation}
for any pair of functions $u_1$, $u_2$ such that $\int_{\rz} u_1^2 \, dx= 
\int_{\rz} u_2^2 \, dx=1$, and $\int_{\rz} u_1 \, u_2 \, dx=0$ 
(i.e., for any pair of mutually orthogonal,
normalized functions). For then, it would follow from (\ref{eq:3.3}) 
and (\ref{eq:3.4}) that
\begin{equation}
\lambda_1+\lambda_2 \le L_{1,1}^1 \int_{\rz} V^{3/2} \, dx.
\label{eq:3.5}
\end{equation}
To prove (\ref{eq:3.4}) is still an open problem. Here we will show that 
(\ref{eq:3.4}) is equivalent to an (open) isoperimetric inequality for 
ovals on the plane. To establish this connection, we perform a change of 
variables similar to the one used in the previous section to prove 
Keller's result on the lowest eigenvalue. 
First we change the independent variable
\begin{equation}
x \to s \equiv \pi \int_{-\infty}^{x} \left(u_1^2+u_2^2 \right) \, dy.
\label{eq:3.6}
\end{equation}
Since $u_1$ and $u_2$ are both normalized, it follows that $s$ runs from 
$0$ to $2 \pi$. 
{}From (\ref{eq:3.6}) we have
$$
\frac{ds}{dx} = \pi \left(u_1^2+u_2^2 \right).
$$
Moreover, set
\begin{equation}
u_1=\rho \cos \theta, \qquad \mbox{and} \qquad u_2=\rho \sin \theta,
\label{eq:3.7}
\end{equation}
so that
\begin{equation}
u_1^2+u_2^2= \rho^2, \qquad \mbox{and} \qquad 
{u_1'}^2+{u_2'}^2= {\rho'}^2 + \rho^2 {\theta'}^2.
\label{eq:3.8}
\end{equation}
With this change of variables we can write
\begin{equation}
\int_{\rz}\left({u_1'}^2+{u_2'}^2 \right) \, dx =
\pi \int_{0}^{2 \pi}
\left(\rho^2 {\dot \rho}^2 + \rho^4 {\dot \theta}^2 \right) \, ds,
\label{eq:3.9}
\end{equation}
and 
\begin{equation}
\int_{\rz}\left(u_1^2+u_2^2 \right)^3 \, dx =
\frac{1}{\pi} \int_{0}^{2 \pi} \rho^4 \, ds. 
\label{eq:3.10}
\end{equation}
Furthermore, set
$$
R=\rho^2,
$$
and
$$
\varphi= 2\theta. 
$$
In these new variables, the desired inequality (\ref{eq:3.4}) is equivalent to
\begin{equation}
\frac
{\int_0^{2\pi} \left({\dot R}^2+R^2{\dot \varphi}^2 \right) \, ds}
{\int_0^{2\pi} R^2\, ds}
\ge 1,
\label{eq:3.11}
\end{equation}
subject to the fact that $u_1$ and $u_2$ are orthonormal, fact that we ought
to express in terms of the new variables. In the new variables,
$$
0=\int_{\rz} u_1 \, u_2 \, dx = \frac{1}{2 \pi}\int_0^{2 \pi} \sin 
\varphi (s) \, ds.
$$
Concerning the other side constraints (i.e., the fact that $u_1$ and 
$u_2$ are normalized), 
given the definition of $s$ and the fact that $s$ runs from $0$ to $2\pi$, 
it is enough to consider the combination
$$
0=\int_{\rz} (u_1^2-u_2^2) \, dx = \frac{1}{\pi}\int_0^{2 \pi} 
\cos \varphi (s) \, ds.
$$
Thus, in the new variables, the fact that $u_1$ and $u_2$ are orthonormal 
imply
\begin{equation}
\int_0^{2 \pi} \sin \varphi (s) \, ds= \int_0^{2 \pi} \cos \varphi (s) \, ds=0.
\label{eq:3.12}
\end{equation}
These latter conditions can be given a simple geometrical interpretation. 
If one considers a closed curve in $\rz^2$ and denote by 
$\cos \varphi (s)$ and 
$\sin \varphi (s)$ the components of the unit tangent, with respect to a fixed
frame, as a function of arc--length, (\ref{eq:3.12}) just says that the curve 
in question is closed. 
Moreover, the curvature of the curve is given by
\begin{equation}
\kappa(s)=\frac{d \varphi}{ds}.
\label{eq:3.13}
\end{equation}
Let's denote by $C$ a closed curve in the plane, of length 
$2 \pi$, with positive curvature, and let
\begin{equation}
H(C) \equiv -\frac{d^2}{ds^2} + \kappa^2
\label{eq:3.14}
\end{equation}
acting on $L^2(C)$ with periodic boundary conditions. Then, (\ref{eq:3.4}), 
and for that matter (\ref{eq:3.11}), is equivalent to saying that the 
lowest eigenvalue of $H(C)$, $\lambda_1(C)$ say,  is
larger or equal to $1$, for any closed curve on the plane of length $2 \pi$. 
It is a simple fact to see that if $C$ is a circle
of length $2 \pi$, the lowest eigenvalue of $H(C)$ is precisely $1$.
Unfortunately we are far from proving the desired bound for general curves.
It is relatively simple to show that the lowest eigenvalue of the Hamiltonian
$H(C)$ is bounded below by $1/2$. To see this one first notes that 
the corresponding eigenfunction can be chosen to be positive. The
quadratic form
$$
(f, H(C) f) = \int_0^{2 \pi} |f'(s)|^2 ds + 
\int _0^{2 \pi} \kappa^2(s) f(s)^2 ds
$$
can be written as 
$$
\int_0^{2 \pi} | \frac{d}{ds} (e^{i \varphi(s)} f(s))|^2 ds \ ,
$$
which we have to minimize over non negative functions $f$ satisfying 
$\int_0^{2 \pi} f(s)^2 ds =1$. 
Expanding the function $e^{i\varphi(s)} f(s)$ into a Fourier series 
$$
e^{i\varphi(s)} f(s) = \sum_{n=-\infty}^\infty c_n 
\frac{ e^{ins}}{\sqrt{2 \pi}} \ ,
$$
we find that since $f(s) \geq 0$
$$
|c_0|^2 \leq \frac{1}{2 \pi} (\int_0^{2 \pi} f(s) ds)^2 \ .
$$
Moreover, since the functions $1/\sqrt{2 \pi}$ and 
$e^{i \varphi(s)}/\sqrt{2 \pi}$ are
orthogonal in the innerproduct of $L^2([0,2 \pi])$ we find that
$$
|c_0|^2 + \frac{1}{2 \pi} (\int_0^{2 \pi}f(s) ds)^2 \leq 
\int_0^{2\pi} f(s)^2 ds =1 \ .
$$
Thus,
$$
|c_0|^2 \leq \min\{\frac{1}{2 \pi} (\int_0^{2 \pi} f(s) ds)^2, 
1-\frac{1}{2 \pi} (\int_0^{2 \pi} f(s) ds)^2\} \leq 1/2 \ .
$$
Since $\sum_n |c_n|^2 =1$ we learn that
$$
\sum_{n \not= 0} |c_n|^2 \geq \frac{1}{2} \ .
$$
Clearly, 
$$
(f, H(C) f) = \sum_{n=-\infty}^{\infty} n^2 |c_n|^2
\geq \sum_{n \not= 0}  |c_n|^2 \geq \frac{1}{2}  \ , 
$$ 
hence $\lambda_1 (C) \ge 1/2$.

\bigskip

\noindent
{\it Remarks:}

\bigskip

\noindent
i) A word of warning should be made at this point. In principle,
the function $R$ defined from the eigenfunctions $u_1$ and $u_2$,
via $\rho$ through equation (\ref{eq:3.8}) above, must vanish at 
$s=0$ and $s=2 \pi$. For the {\it curve problem}, however, we drop
this boundary condition. Thus, a priori the conjecture for the 
{\it curve problem} is stronger than the Lieb--Thirring conjecture
for the two bound states, although we believe it amounts to the same.

\bigskip

\noindent
ii) The best bound to date on $L_{1,1}$ is the bound of Eden and Foias 
\cite{EdFo91}
 who proved,
\begin{equation}
L_{1,1} \le \frac{2}{9} \sqrt{3} \approx 0.3849 \dots
\label{eq:ed-fo-bound}
\end{equation}
Our bound $\lambda_1(C) \ge 1/2$ yields the bound
$$
L_{1,1} \le \frac{4}{9 \pi} \sqrt{6} \approx 0.3465 \dots,
$$
which although  better than (\ref{eq:ed-fo-bound}),  only 
applies to Schr\"odinger operators with two bound states. Just for 
comparison, the conjectured sharp value for $L_{1,1}$ is 
$4\sqrt{3}/(9 \pi) \approx 0,2450$.

\bigskip

\noindent
iii) In recent years several authors have obtained isoperimetric inequalities
for the lowest eigenvalues of a variant of $H(C)$, and we give a short 
summary of the main results in the sequel. Consider the Schr\"odinger operator
\begin{equation}
H_g(C) \equiv -\frac{d^2}{ds^2} + g \kappa^2
\label{eq:3.15}
\end{equation}
defined on $L^2(C)$ with periodic boundary conditions. As before, $C$ denotes
a closed curve in $\rz^2$ with positive curvature $\kappa$, and length $2 \pi$.
Here, $s$ denotes arclength. If $g<0$, the lowest eigenvalue of $H_g(C)$, say
$\lambda_1(g,C)$ is uniquely maximized when $C$ is a circle \cite{DuEx95}. 
When $g=-1$, the second eigenvalue, $\lambda_2(-1,C)$ is uniquely maximized 
when $C$ is a circle \cite{HaLo98}. If $0,g \le 1/4$, 
$\lambda_1(g,C)$ is uniquely 
minimized when $C$ is a circle \cite{ExHaLo99}. It is an open problem to 
determine
the curve $C$ that minimizes $\lambda_1(g,C)$ in the cases, 
$1/4 < g \le 1$, and
$g<0, g \neq -1$. If $g > 1$ the circle is not a minimizer for 
$\lambda_1(g,C)$
(see, e.g., \cite{ExHaLo99,Ha02} for more details on the subject). 

\bigskip
\bigskip

To conclude this section we give an alternative interpretation of the 
minimization principle (\ref{eq:3.11}) subject to the side constraints 
(\ref{eq:3.12}). Interpret now $s$ as time (instead of arclength) and, 
given $R(s)$ and $\varphi(s)$ as before, 
define
$$
x(s) = R(s) \cos \varphi (s),
$$
and 
$$
y(s) = R(s) \sin \varphi (s).
$$
Then, the minimization problem (\ref{eq:3.11}), (\ref{eq:3.12}) is 
equivalent to the following,
\begin{equation}
\frac{\int_0^{2 \pi} \left( {\dot x}^2 + {\dot y}^2 \right) \, ds}
{\int_0^{2 \pi} \left( x^2 + y^2 \right) \, ds} \ge 1,
\label{eq:3.16}
\end{equation}
where $x(s)$ and $y(s)$ are periodic, of period $2 \pi$ and satisfy the
side constraints,
\begin{equation}
\int_0^{2 \pi} \frac{x(s)}{\sqrt{x(s)^2+y(s)^2}} \, ds=
\int_0^{2 \pi} \frac{y(s)}{\sqrt{x(s)^2+y(s)^2}} \, ds= 0.
\label{eq:3.17}
\end{equation}
Notice that (\ref{eq:3.16}) certainly holds if one replaces 
the side constraints (\ref{eq:3.17}) by $\int_0^{2 \pi} x(s) \, ds =0$ 
and $\int_0^{2 \pi} y(s) \, ds =0$, for then both functions $x(s)$ and
$y(s)$ would be orthogonal to the constants and one would have
$\int_0^{2 \pi} {\dot x}^2 \, ds \ge \int_0^{2 \pi} x(s)^2 \, ds$ and 
$\int_0^{2 \pi} {\dot y}^2 \, ds \ge \int_0^{2 \pi} y(s)^2 \, ds$,
independently.

\section{Appendix}

To obtain inequality (\ref{eq:2.2}) for $\gamma \neq 1$ we start from 
equation (\ref{eq:2.5}) as before. Using H\"older's inequality we get
\begin{equation}
\lambda_1 \le 
\left(
\int_{-\infty}^{\infty} V^{\gamma+(1/2)} \, dx
\right)^{2/2\gamma+1}
\left(\int_{-\infty}^{\infty}u_1^{2 (2\gamma+1)/(2 \gamma-1)} \, dx
\right)^{(2\gamma-1)/(2\gamma+1)} 
- \int_{-\infty}^{\infty} (u_1')^2 \, dx.
\label{eq:a0}
\end{equation}
We claim that if  $\int_{-\infty}^{\infty} u_1^2 \, dx=1$, 
\begin{equation}
\int_{-\infty}^{\infty} {u_1'}^2 \, dx \ge c(\gamma) 
\left( \int_{-\infty}^{\infty} 
u_1^{2(2\gamma+1)/(2\gamma-1)} \,  dx \right)^{2 \gamma-1},
\label{eq:a1}
\end{equation}
where
$$
c(\gamma)= 
\left[\sqrt{\frac{\pi}{2}} \frac{\gamma^{\gamma} \Gamma(\gamma+1/2)}
{\Gamma(\gamma+1)(\gamma-1/2)^{\gamma-1/2}}
\right]^2.
$$
Using the claim and denoting
$$
A \equiv \left( \int_{-\infty}^{\infty} V^{\gamma + (1/2)}\, dx \right)^{2/2\gamma+1},
$$
and 
$$
Y \equiv  \left( \int_{-\infty}^{\infty} 
u_1^{2(2\gamma+1)/(2\gamma-1)} \, dx \right)^{2 \gamma-1},
$$
we get
\begin{equation}
\lambda_1 \le A Y^{1/(2\gamma+1)} - c(\gamma) Y.  
\label{eq:a2}
\end{equation}
Maximizing the left side of (\ref{eq:a2}) over $Y$ (for $\gamma> 1/2$), we get
\begin{equation}
\lambda_1 \le \tilde c(\gamma) \left( \int_{-\infty}^{\infty} 
V^{\gamma + (1/2)}\, dx \right)^{1/\gamma},
\label{eq:a3}
\end{equation}
where
$$
\tilde c (\gamma) = \frac{2 \gamma}{c(\gamma)^{1/(2 \gamma)}
(2\gamma+1)^{(2\gamma+1)/(2 \gamma)}}.
$$
Hence,
\begin{equation}
\lambda_1^{\gamma} \le L_{\gamma,1}^1 \int_{-\infty}^{\infty} 
V^{\gamma + (1/2)}\, dx.
\label{eq:a4}
\end{equation}
To conclude we need only to prove the claim (\ref{eq:a1}) 
whenever $\int_{-\infty}^{\infty}
u_1^2 \, dx=1$. Introducing the same change of variables as in Section 2, 
i.e., 
$$
x \to s = \int_{-\infty}^{x} u_1^2 \, dy,
$$
and 
$$
w\equiv u_1^2, 
$$
the claim reduces to proving
\begin{equation}
\frac{1}{4} \int_0^1 {\dot w}^2 \, ds \ge c(\gamma) \left(
\int_0^1 w^{2/2\gamma-1} \, ds \right)^{2 \gamma -1},
\label{eq:a5}
\end{equation}
which follows from Sobolev's inequality in one dimension.

\section{Acknowledgements}

It is a pleasure to thank the organizers of the Pan-American Advanced 
Studies Institute (PASI) on Partial Differential Equations, 
Inverse Problems and Non-Linear Analysis for their kind
invitation to present these results. 
We thank Mark S. Ashbaugh, Evans Harrell and  Elliott
Lieb for useful discussions.

\bibliographystyle{amsalpha}

\end{document}